\PassOptionsToPackage{table}{xcolor}
\documentclass[APA,STIX1COL]{WileyNJD-v2}

\newcommand\BibTeX{{\rmfamily B\kern-.05em \textsc{i\kern-.025em b}\kern-.08em
T\kern-.1667em\lower.7ex\hbox{E}\kern-.125emX}}

\usepackage{rotating}

\usepackage{graphicx}

\usepackage[english]{babel}

\usepackage{caption}
\usepackage{subcaption}

\usepackage[]{mdframed}

\usepackage{ulem}

\usepackage{orcidlink}

\articletype{Systematic Review}%

\begin{document}

\title{A Systematic Review on Process Mining for Curricular Analysis}

\author[1,2]{Daniel Calegari}

\author[2]{Andrea Delgado}

\authormark{Daniel Calegari and Andrea Delgado}

\address[1]{\orgname{Universidad ORT Uruguay}, \orgaddress{\state{Montevideo, 11100}, \country{Uruguay}}}

\address[2]{\orgdiv{Instituto de Computación}, \orgname{Facultad de Ingeniería, Universidad de la República}, \orgaddress{\state{Montevideo, 11300}, \country{Uruguay}}}

\corres{Daniel Calegari, \email{calegari@ort.edu.uy}}


\abstract[Abstract]{Educational Process Mining (EPM) is a data analysis technique that is used to improve educational processes. It is based on Process Mining (PM), which involves gathering records (logs) of events to discover process models and analyze the data from a process-centric perspective. One specific application of EPM is curriculum mining, which focuses on understanding the learning program students follow to achieve educational goals. This is important for institutional curriculum decision-making and quality improvement. Therefore, academic institutions can benefit from organizing the existing techniques, capabilities, and limitations.
We conducted a systematic literature review to identify works on applying PM to curricular analysis and provide insights for further research. 
We reviewed 27 primary studies published across seven major databases. Our analysis classified these studies into five main research objectives: discovery of educational trajectories, identification of deviations in student behavior, bottleneck analysis, dropout / stopout analysis, and generation of recommendations. Key findings highlight challenges such as standardization to facilitate cross-university analysis, better integration of process and data mining techniques, and improved tools for educational stakeholders. This review provides a comprehensive overview of the current landscape in curricular process mining and outlines specific research opportunities to support more robust and actionable curricular analyses in educational settings.
}


\maketitle

\section{Introduction}\label{sec:introduction}

Educational Data Mining (EDM) \citep{SLRAux2,CA4_article} is a field that deals with the analysis of data from educational information systems such as Learning Management Systems (LMS) and administrative systems. Its primary objective is to answer critical academic questions and improve educational practices. To achieve this, EDM employs different data mining techniques to gain insights into the educational context \citep{CA4_article}, e.g., grouping similar students based on their learning and interaction patterns and detecting students with difficulties or irregular learning processes, and also to intervene in such context, e.g., predicting an outcome and recommend a more adequate strategy.

\begin{figure}
\centering
\begin{subfigure}{.5\textwidth}
  \centering
  \includegraphics[width=\linewidth]{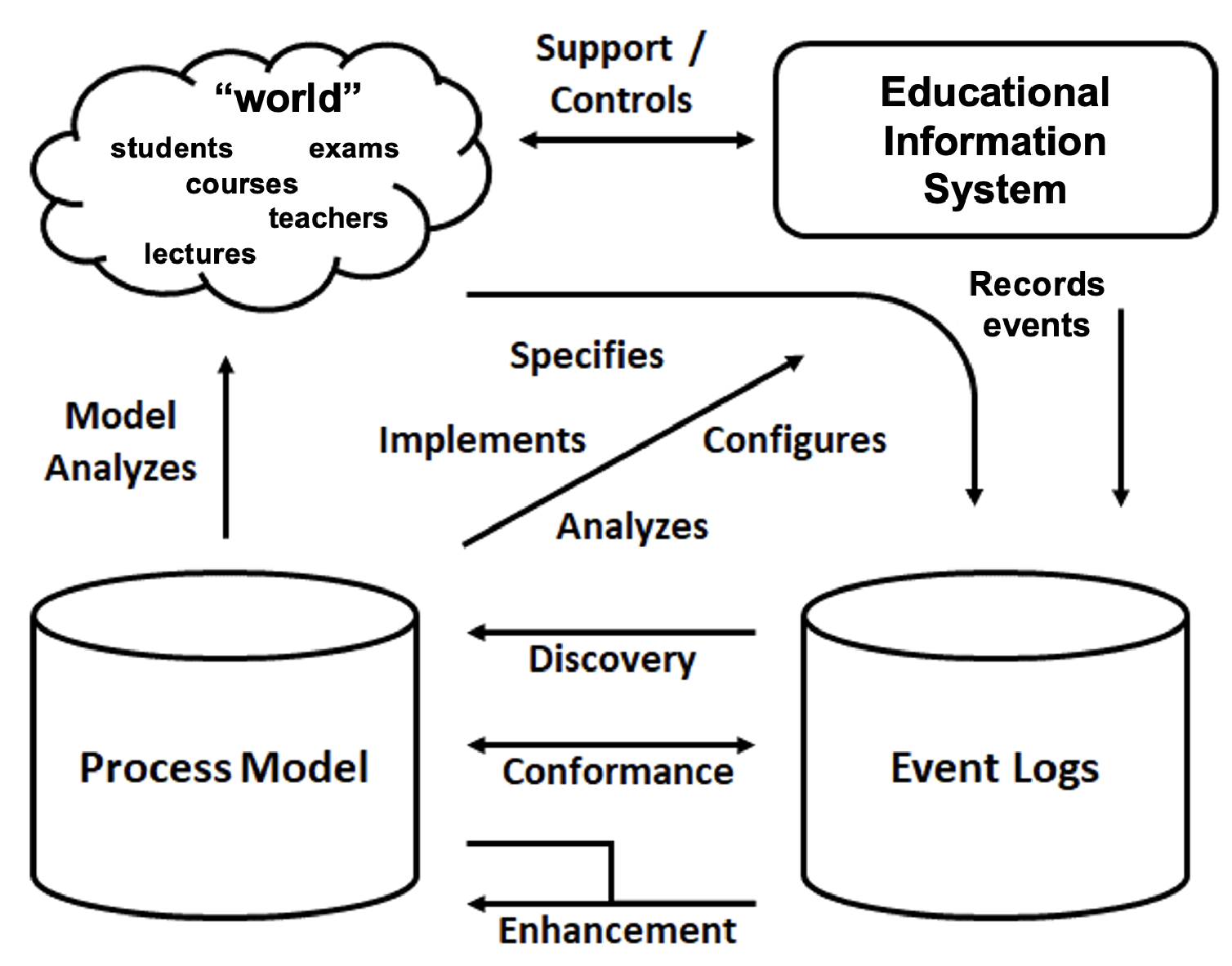}
\end{subfigure}%
\begin{subfigure}{.5\textwidth}
  \centering
  \includegraphics[width=\linewidth]{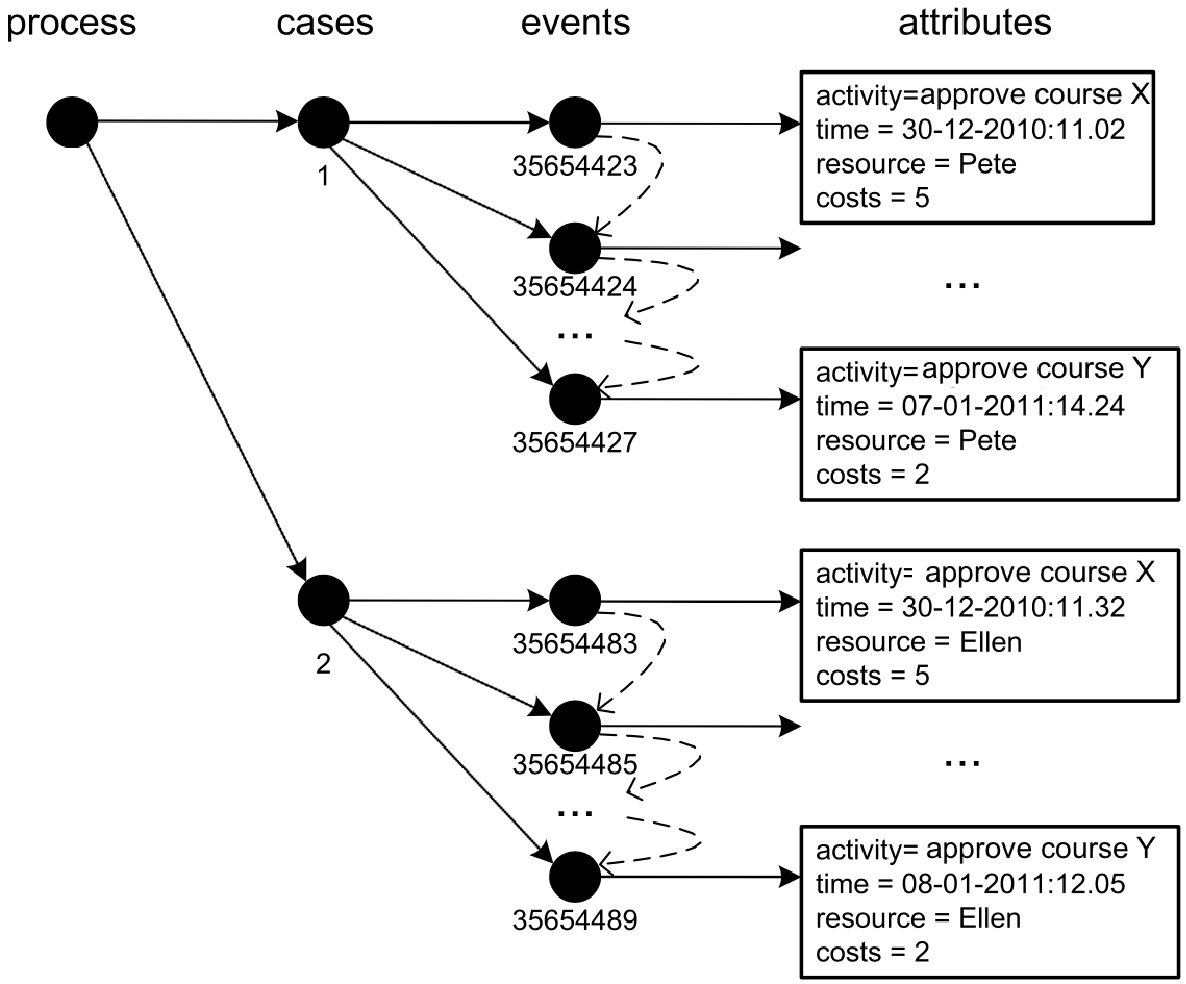}
\end{subfigure}
\caption{EPM schema and structure of an event log (both adapted from \citep{ProcessMiningBook})}
\label{fig:epmschema}
\end{figure}

This data-centric view of educational data is refined by the process-centric view provided by Educational Process Mining (EPM) \citep{PMEducation}. 
In general terms, Process Mining (PM) \citep{ProcessMiningBook} allows analyzing the records (logs) of events associated with the execution of educational processes, as shown in Figure \ref{fig:epmschema}. The analysis involves three types of activities: \textbf{discovery}, which uses event logs to create a process model from the causal dependencies between events; \textbf{conformance}, which implies verifying the correspondence of the enacted processes concerning the expected one (through a reference model); and \textbf{enhancement}, which uses additional information to extend or improve an existing process, e.g., adding new perspectives about the duration of the process, bottlenecks or the underuse of resources, among others. 

Event logs are extracted from educational information systems and are traditionally organized according to the following principles \citep{ProcessMiningBook}:
\begin{itemize}
    \item A process comprises \textbf{cases} (traces) that represent the execution of a process instance from start to finish. Each case has a case ID, allowing one to identify a case among others uniquely. As an example, shown in Figure \ref{fig:epmschema}, it could be possible to consider as a case the approval history of a student, with the student ID identifying each case.
    \item A case comprises \textbf{events}, i.e., an action performed in the process. Each event is associated with precisely one case. In the example, an event is the approval of a course.
    \item Events can have activity, time, cost, and resource \textbf{attributes}. At least one attribute must be present that represents the activity carried out, i.e., its activity name. In the example, the course name can be taken as the activity name, and the approval date, student, and credits earned are additional attributes of such an event.
    \item Events within a \textbf{case are ordered} (usually by a timestamp), determining a partial order between events of the same case. In the example, the approval date defines an ordering of the events.
\end{itemize}

Depending on the definition of a case and its events, there are different perspectives from which the data can be analyzed. As synthesized in previous systematic literature reviews \citep{SLR2,SLR1,SLR3}, EPM pursues different objectives, e.g., to gain a better understanding of the underlying educational process, to detect collaborative or self-regulated student's learning behavior and difficulties, and to generate recommendations or advice for students and teachers, among others. Many of these objectives relate to analyzing learner's behavior in the context of a course, e.g., gathering data from Learning Management Systems (LMS) usage based on actual learner's activity recorded in the event logs \citep{SLRAux1}. However, stakeholders such as study program coordinators and advisors must consider the institutional viewpoint for curriculum understanding and improvement. 

An academic curriculum defines a specific learning program \citep{4_Mykola}. It constrains how students are expected to take the courses \citep{8_Wang}, e.g., defining that a particular course is compulsory or must be taken before or after another. Depending on the flexibility of the curriculum, students have more or less freedom to attend courses and take exams in different order. Curricular improvements must consider students' behaviors while trying to achieve the desired educational goals, such as a low dropout rate and a high percentage of students who graduate on time \citep{18_Priyambada}.

Curriculum mining/analytics \citep{CA1_Hilliger} focuses on collecting, analyzing, and visualizing the program- and course-level data, such as program structure and course grading, to support curriculum decision-making and quality improvement specifically. It allows analyzing the strengths and weaknesses of a curriculum by identifying its components, relations, and constraints, evaluating how they fit together, identifying potential and actual problems (e.g., blind spots, bottlenecks, and deviations of students' trajectories), checking underlying beliefs and assumptions, and determining whether the goals have been met.

Curriculum mining is identified as an application domain of EPM \citep{SLR1}. There are general reviews on EDM \citep{SLRAux2}, EPM \citep{SLR2,SLR1,SLR3}, and PM in LMS \citep{SLRAux1} that mentioned curricular mining. 
While numerous studies have applied EPM techniques to various educational contexts, the specific domain of curricular mining lacks a comprehensive synthesis. This gap hinders the development of standardized methods and limits the scalability of insights across institutions. 


In this work, we conduct a Systematic Literature Review (SLR) following the methodological guidelines in \citep{Kit04, kit07} to identify works on applying PM to curricular analysis. We also classify and analyze the existing work to evaluate the PM scope in this domain and provide insights for further research. 
This paper contributes to the field by: (1) synthesizing findings from 27 studies on process mining applications in curricular analysis, (2) identifying five primary objectives pursued in this domain, and (3) proposing a research agenda to address open challenges such as standardization, tool integration, and the alignment of process mining with data mining techniques.

The rest of the paper is organized as follows. 
In Section \ref{sec:method}, we describe the conducted literature review methodology, and in Section \ref{sec:results}, we synthesize the findings. 
Then, in Section \ref{sec:discussion}, we discuss the findings and investigate future research directions. 
In Section \ref{sec:threats}, we present the threats to the validity of our review.
Finally, in Section \ref{sec:conclusions}, we provide the conclusions of this work.
\section{Literature review methodology}\label{sec:method}

\begin{figure}[!ht]
    \centering
    \includegraphics[width=0.8\linewidth]{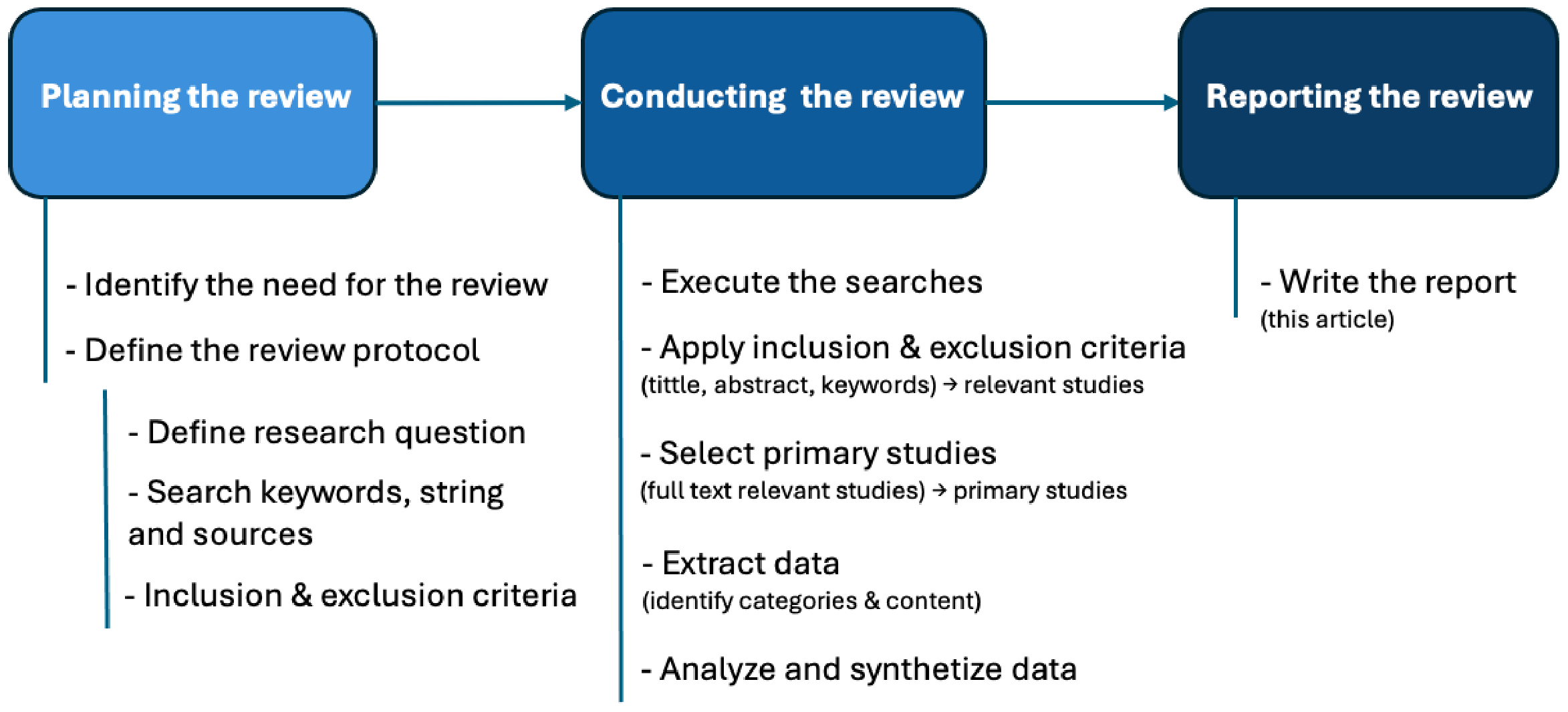}
    \caption{SLR research methodology summary}
    \label{img:slrSummary}
\end{figure}

We conducted a Systematic Literature Review (SLR) based on \citep{Kit04, kit07}. As depicted in Figure \ref{img:slrSummary}, it defined three phases: planning, conducting, and reporting the review, summarized next.  

\subsection{Planning the review}

This phase identifies the need for a systematic review and defines the protocol for conducting it. The protocol includes formulating the research question, describing the search terms and string, selecting the search sources to be used, and defining the inclusion and exclusion criteria for identifying the primary studies of the review.

In this work, the need for a systematic review is identified in the context of EDM and EPM initiatives we were carrying out within the Computer Science career of the Instituto de Computación (Computer Science Institute, InCo) of the Facultad de Ingeniería (Engineering School, FING) of the Universidad de la República (UdelaR), to be used as the basis for the career analysis and students paths towards the defined curricula. The result of this review allows us to identify critical elements to be considered in such analysis. It also made us realize that the lack of standardization between studies makes it very hard to compare curricula of the same career from different countries, even from other universities of the same country. 

This work focuses on EPM initiatives, for which we defined the review protocol presented next.\\

\noindent\textbf{Define research question}
The general research question guiding the review is 

\begin{mdframed}
\begin{center}
\textit{\textbf{What are the research papers on applying process mining to curricular analysis ?}} 
\end{center}
\end{mdframed}

Since it involves an understanding of the objectives pursued by related articles and the techniques they used, as well as identifying possible open research opportunities, we derived specific questions to drive the analysis:

\begin{description}
\setlength{\itemindent}{.3in}
    \item[\textbf{(RQ1)}] What are the research objectives of these works?
    \item[\textbf{(RQ2)}] How PM techniques are used?
    \item[\textbf{(RQ3)}] What research methodologies and PM methods and tools are used?
    \item[\textbf{(RQ4)}] What are the open challenges and research opportunities?
\end{description}

\noindent\textbf{Search keywords, strings, and sources}
We defined a search string for the selection of published papers. We search for keywords in existing literature reviews \citep{SLR1,SLR2} and in some papers referred by them. Since many heterogeneous keywords arise, e.g., curriculum, dropout, bottleneck, registration, and trajectories, we used the general string
\begin{center}
\textsl{``process mining''} \textsl{AND} \textsl{``education*''}
\end{center}
which considers education, educational, and education data. These keywords lead to more false positive matches, but they help cover a broader spectrum of the literature. False positives are filtered out using the inclusion and exclusion criteria introduced in the next step and the manual inspection of articles.
We identified IEEE Xplore, ACM Digital Library, SpringerLink, ScienceDirect, Wiley, Scopus, and Web of Science as sources.\\

\noindent\textbf{Inclusion and exclusion criteria} 
Table \ref{tab:criteria} shows the inclusion and exclusion criteria we have considered. Since we are interested in extracting all kinds of ideas, even those still incipient, we did not exclude short papers. Moreover, we consider extensions of previous articles to have a complete view of the works.

{\setlength{\tabcolsep}{4pt}
\begin{table}[!ht]
\centering
\caption{Inclusion and exclusion criteria.}
\label{tab:criteria}
\begin{tabular}{p{7cm} p{10cm}} 
 \hline
 \textbf{Inclusion} & \textbf{Exclusion} \\
 \hline
 (1) The paper concerns curricular analysis.
 & 
 (1) The study is not written in English.
 \\
 (2) The study employs a PM approach as a central element of its analysis.
 &
 (2) The paper refers to some non-peer-reviewed publications, such as technical reports, book chapters, proceedings’ prefaces, and editorials. 
 \\
 &
 (3) The paper is not electronically available on the web.
 \\
 \hline
\end{tabular}
\end{table}
}

\subsection{Conducting the review}

Based on the defined protocol, we conducted the review following the steps already summarized in Figure \ref{img:slrSummary}. In what follows, we describe the first four steps. The last step concerning the data analysis is thoughtfully described in Section \ref{sec:results} and Section \ref{sec:discussion}. \\

\noindent\textbf{Execute the searches}
The initial search covered papers until 2th November 2023.
We then extended the search covering papers until 7th November 2024, giving a total of 3.481 articles, as shown in Table \ref{tab:search}.\\

\noindent\textbf{Apply inclusion \& exclusion criteria}
We applied inclusion and exclusion criteria (Table \ref{tab:criteria}) by reading the title, abstract, and keywords of each paper. This refinement gives a primary selection of 90 papers (with duplicates), as shown in Table \ref{tab:search}.\\

\noindent\textbf{Select primary studies}
We selected the primary studies based on reading the full text of the paper. We reinforced the results by backward searching references (snowballing), searching for authors’ and conferences’ web pages, Google Scholar, and the DBLP database. It allowed finding other papers marked as ``Other'' in Table \ref{tab:search}, giving 43 papers. After removing duplicates, we finally identified 27 primary studies. In the last column of Table \ref{tab:search}, we keep one of the sources for duplicates (ScienceDirect, Scopus and Web of Science are the sources that have duplicates already in the other sources). The primary studies we selected deal with the specific curricular analysis perspective we want to analyze, which was only partially introduced in three existing literature reviews that partially refer to the topic \citep{SLR2,SLR1,SLR3}.\\

\noindent\textbf{Extract data} 
We extracted the primary data of each work summarized in Table \ref{tab:primaryStudiesReduced} (ordered by year/author). We also extracted data that allowed us to categorize the works and answer the research questions that will be analyzed in the following sections. 

{\setlength{\tabcolsep}{4pt}
\begin{table}[!ht]
\centering
\caption{Data sources and results for literature search.}
\label{tab:search}
\begin{tabular}{l r r r r} 
 \hline
 \textbf{Source} & \textbf{Total results} & \textbf{Relevant studies} & \textbf{Primary studies w/duplicates} & \textbf{Primary studies} \\
 \hline
 IEEE Xplore & 67 & 17 & 5 & 5\\
 \hline
 ACM Digital Library & 369 & 3 & 2 & 2\\
 \hline
 SpringerLink & 2428 & 14 & 7 & 7\\
 \hline
 ScienceDirect & 20 & 3 & 1 & 0\\
 \hline
 Wiley & 165 & 7 & 0 & 0\\
 \hline
 Scopus & 295 & 36 & 14 & 7\\
 \hline
 Web of Science & 137 & 10 & 8 & 0\\
 \hline
 Other & -- & -- & 6 & 6 \\
 \hline
 \textbf{Total} & \textbf{3481} & \textbf{90} & \textbf{43} & \textbf{27}\\
 \hline
\end{tabular}
\end{table}
}

\begin{sidewaystable}
\caption{Primary studies}
\centering
{\scriptsize
\begin{tabular}{|c|c|l|p{8.5cm}|l|}
\hline
\textbf{Ref} & 
\textbf{Year} & 
\textbf{Main author} & 
\textbf{Title} &
\textbf{Source}\\ 
\hline\hline

\citep{1_Trcka} &
2009 &
Trcka N. &
From local patterns to global models: Towards domain driven educational process mining &
IEEE
\\ \hline

\citep{2_Anuwatvisit} &
2012 &
Anuwatvisit S. &
Bottleneck mining and Petri net simulation in education situations  &
IEEE
\\ \hline

\citep{3_Ayutaya} &
2012 &
Ayutaya N. &
Heuristic mining: Adaptive process simplification in education  &
IEEE
\\ \hline

\citep{4_Mykola} &
2012 &
Pechenizkiy M.&
CurriM: Curriculum mining &
Scopus 
\\ \hline

\citep{24_Awatef} &
2014 &
Cairns A.&
Towards Custom-Designed Professional Training Contents and Curriculums through Educational Process Mining &
Other
\\ \hline

\citep{7_Awatef} &
2014 &
Cairns A.&
Using semantic lifting for improving educational process models discovery and analysis  &
Scopus
\\ \hline

\citep{25_Bendatu_Yahya_2015} &
2015 &
Bendatu L.&
Sequence Matching Analysis for Curriculum Development  &
Other
\\ \hline

\citep{8_Wang} &
2015 &
Wang R.&
Discovering process in curriculum data to provide recommendation  &
Other
\\ \hline

\citep{9_Cameranesi} &
2017 &
Cameranesi M.&
Students' careers analysis: A process mining approach  &
Scopus
\\ \hline

\citep{11_Schulte} &
2017 &
Schulte J.&
Large scale predictive process mining and analytics of university degree course data  &
Scopus
\\ \hline

\citep{12_Caballero} &
2018 &
Caballero J.&
Discovering bottlenecks in a computer science degree through process mining techniques  &
IEEE
\\ \hline

\citep{13_Wang} &
2018 &
Wang R.&
Sequence-based Approaches to Course Recommender Systems  &
Springer
\\ \hline

\citep{14_Salazar} &
2019 &
Salazar J.&
Describing educational trajectories of engineering students in individual high-failure rate courses that lead to late dropout  &
Scopus
\\ \hline

\citep{15_Salazar} &
2019 &
Salazar J.&
Influence of Student Diversity on Educational Trajectories in Engineering High-Failure Rate Courses that Lead to Late Dropout  &
IEEE
\\ \hline

\citep{17_Martinez} &
2021 &
Martinez P.&
Modelling Computer Engineering Student Trajectories with Process Mining  &
Other
\\ \hline

\citep{19_Salazar} &
2021 &
Salazar J.&
Backpack process model (Bppm): A process mining approach for curricular analytics  &
Scopus
\\ \hline

\citep{20_Salazar} &
2021 &
Salazar J.&
Curricular analytics to characterize educational trajectories in high-failure rate courses that lead to late dropout  &
Scopus
\\ \hline

\citep{23_Hobeck} &
2022 &
Hobeck R.&
Process Mining on Curriculum-Based Study Data: A Case Study at a German University  &
Springer
\\ \hline

\citep{21_mahamed} &
2022 &
Mahammed N.&
A System to Search and Recommend Learning Courses Sequences  &
Other
\\ \hline

\citep{22_Wagner} &
2022 &
Wagner M.&
A Combined Approach of Process Mining and Rule-based AI for Study Planning and Monitoring in Higher Education  &
Springer
\\ \hline

\citep{Potena23} &
2023 &
Potena D.&
Evidence-based Student Career and Performance Analysis with Process Mining: a Case study &
Other
\\ \hline

\citep{Rafied23} &
2023 &
Rafied M.&
Extracting Rules from Event Data for Study Planning &
Other
\\ \hline

\citep{Diamantini24b} &
2024 &
Diamantini C.&
Evidence-driven appraisal of students’ careers using process mining: a case study &
Springer
\\ \hline

\citep{Diamantini24} &
2024 &
Diamantini C.&
Understanding the stumbling blocks of Italian higher education system: A process mining approach &
ACM
\\ \hline

\citep{Puttow24} &
2024 &
Puttow L.&
Towards a labeling method for Education Process Mining and a case study on higher education &
ACM
\\ \hline

\citep{Rennert24} &
2024 &
Rennert C.&
Evaluation of Study Plans using Partial Orders &
Other
\\ \hline

\citep{Roepke24} &
2024 &
Roepke R.&
Study path analyses for quality assurance and support of study planning &
Springer
\\ \hline

\end{tabular}
}
\label{tab:primaryStudiesReduced}
\end{sidewaystable}
\section{Data synthesis \& results}\label{sec:results}

\begin{figure}[!ht]
\centering
\begin{subfigure}{.8\textwidth}
    \centering
    \includegraphics[width=\linewidth]{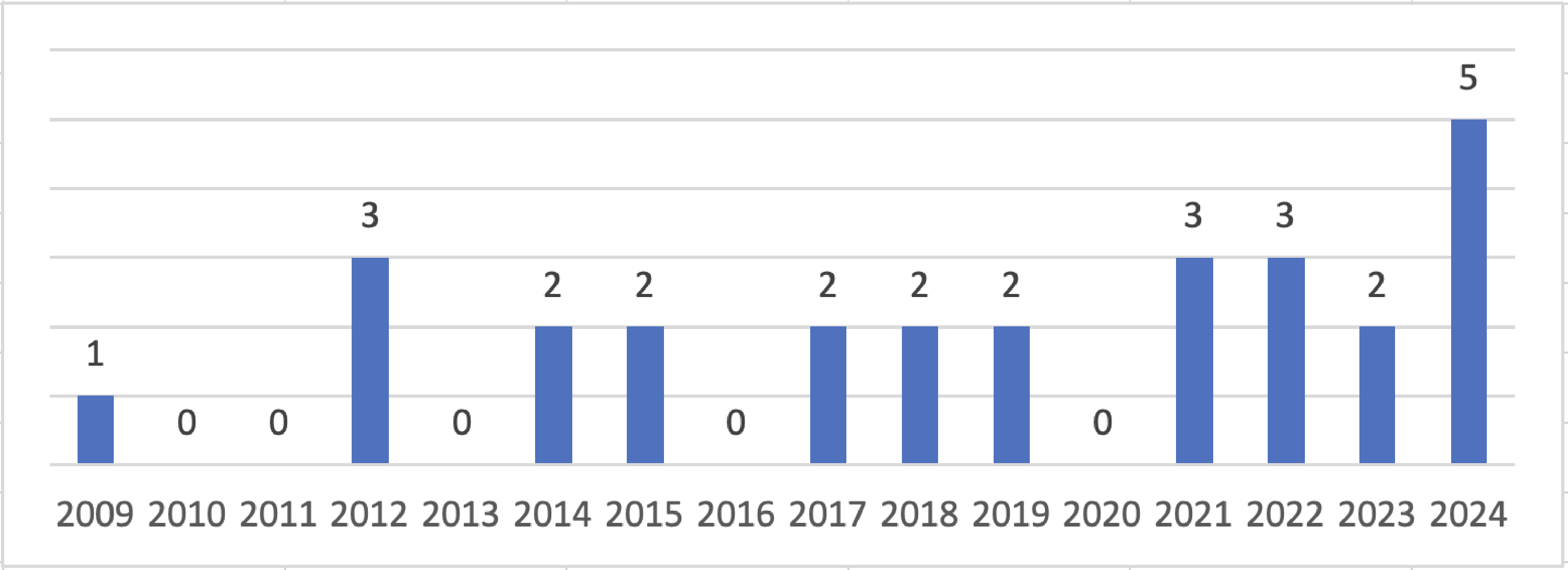}
\end{subfigure}
\vfill
\vspace{0.2cm}
\begin{subfigure}{.6\textwidth}
    \centering  
    \includegraphics[width=\linewidth]{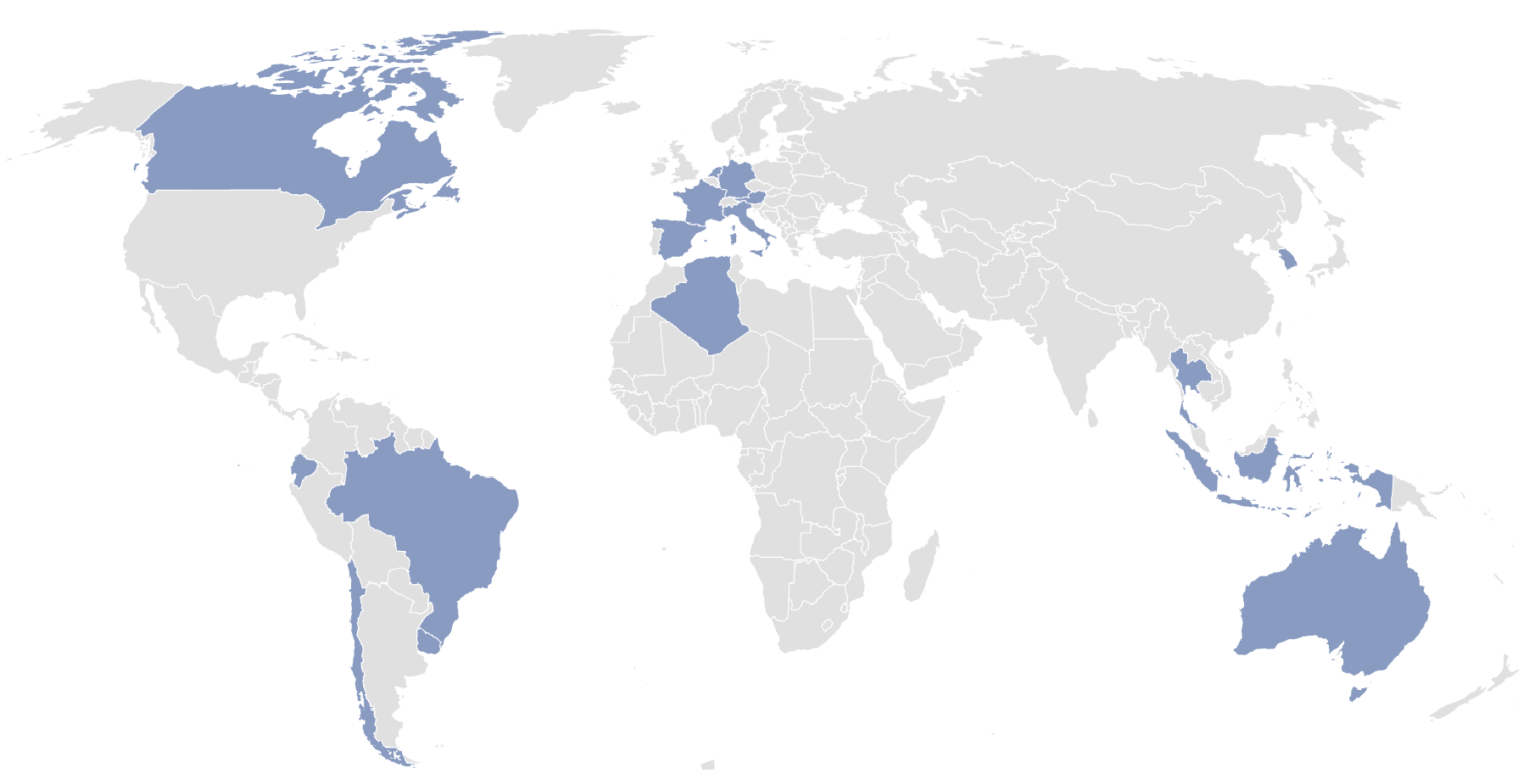}
\end{subfigure}
\begin{subfigure}{.35\textwidth}
    \centering  
    \includegraphics[width=\linewidth]{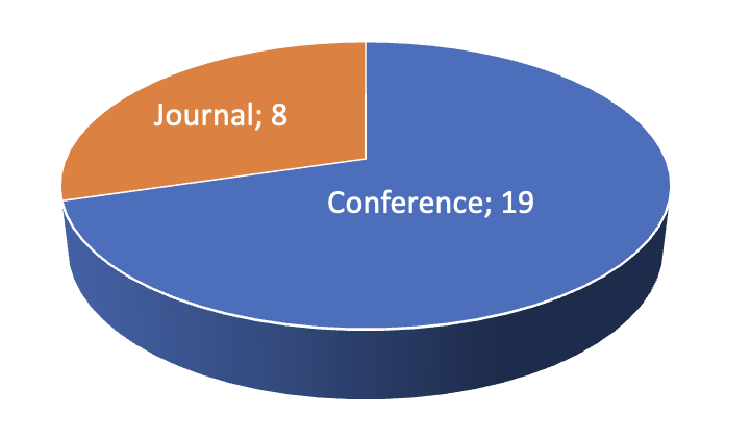}
\end{subfigure}
\caption{Data summary: number of papers published by year, countries of affiliation of researchers, and kind of publications}
\label{fig:datasummary}
\end{figure}

This section synthesizes and analyzes the collected data, answering the first two research questions. Figure \ref{fig:datasummary} summarized primary data; it depicts the distribution of the research papers through the years, the countries of the affiliation of researchers, and the type of publication (conference or journal). 

As can be seen, the topic has been studied regularly for the last ten years. The interest is spread in universities in fourteen countries, including Asia, Europe, Oceania, and South and North America. Works have been published primarily at conferences. In this case, until 2015, papers were published in more general venues. Since 2015, papers have been published in more specific EDM and education venues, such as the International Conference on Educational Data Mining and the International Conference on Learning Analytics \& Knowledge. Moreover, since 2022, papers have been published in a more dedicated venue: the Education meets Process Mining (EduPM) workshop at the International Conference on Process Mining (ICPM). In the case of papers published in journals, even for recent publications from 2022, it can be seen that there are published in non-specialized journals in the EPM, EDM, or the education domains, but on journals focused on more general domains, such as the Database and Expert Systems Applications, Applied Sciences, and the Turkish Journal of Computer and Mathematics Education.

\subsection*{(RQ1)
What are the research objectives of these works?}

The main objective addressed in these papers is \textbf{educational trajectories discovery}. It means discovering a process model that reflects the trajectory of students, considering a trajectory as the set of courses, exams, or curriculum milestones taken or approved by students on a given period. It also involves identifying the many different variants of students' behavior and the students' characteristics that describe such variants. 
The discovery of educational trajectories is strongly related to the formal modeling of an academic curriculum from predefined requirements. This process considers the curriculum's structure, prerequisites, and business rules, such as courses that cannot be taken in parallel, approval deadlines, and the maximum number of courses that can be taken in one semester.

Using a discovered or pre-authored curriculum model, it is possible to perform \textbf{curricular conformance checking} to check whether the observed behavior of students matches their expected behavior to identify deviations. As with the discovery of trajectories, it also involves analyzing variants of students' behavior. This objective could be used to understand how the curriculum's structure changes when certain administrative decisions are taken, e.g., allowing students to take courses without having all the prerequisites.

Even without deviations from the expected curricular trajectory, there could be problems with students' progress. \textbf{Bottleneck analysis} focuses on identifying and analyzing causes delaying progress (bottlenecks) on the student's educational trajectories. The \textbf{stopout/dropout analysis} is a more specific analysis, not focused on delays, but on identifying and analyzing causes that lead to stopout and dropout of students. 
In both cases, predictive analytics can be used to anticipate such problems, and the cause analysis is focused on identifying the courses/exams, students' characteristics, and behavioral patterns that lead to them.

Finally, \textbf{trajectory/curriculum recommendation} is devoted to making personalized recommendations to the student's educational trajectories or the curricular design to improve the expected results, e.g., to earn a degree. These recommendations are informed by the insights gathered through trajectory discovery, conformance checking, bottleneck, and dropout analyses, collectively guiding improvements in individual student success and curriculum effectiveness.

\setlength\extrarowheight{2pt}
\begin{table}[!ht]
\caption{Papers focused on each research objective.}
\centering
{\footnotesize
\begin{tabular}{|l|c|l|}
\hline
\textbf{Objective} & 
\textbf{\#} & 
\textbf{References} \\ 
\hline\hline

\multirow{4}{8cm}{\underline{\textbf{Educational trajectories discovery}}: discovering a process model that reflects the educational trajectory of students.} & 
17 &
\citep{1_Trcka} \citep{2_Anuwatvisit} \citep{3_Ayutaya} \\ 
&&
\citep{4_Mykola} \citep{7_Awatef} \citep{24_Awatef} \\
&&
\citep{8_Wang} \citep{9_Cameranesi} \citep{11_Schulte} \\
&&
\citep{12_Caballero} \citep{17_Martinez}\\
&&
\citep{Potena23} \citep{23_Hobeck} 
\\ 
&&
\citep{Diamantini24b} \citep{Diamantini24} 
\\ 
&&
\citep{Puttow24} \citep{Roepke24} 
\\ 
\hline

\multirow{3}{8cm}{\underline{\textbf{Curricular conformance checking}}: checking whether the observed students' educational trajectories match their expected curricula.} & 
14 &
\citep{1_Trcka} \citep{3_Ayutaya} \citep{4_Mykola} \\ 
&&
\citep{7_Awatef} \citep{8_Wang} \citep{25_Bendatu_Yahya_2015}\\
&&
\citep{9_Cameranesi} \citep{22_Wagner}\\
&&
\citep{Potena23} \citep{23_Hobeck}
\\ 
&&
\citep{Diamantini24b} \citep{Diamantini24} 
\\ 
&&
\citep{Rennert24} \citep{Roepke24} 
\\ \hline

\multirow{2}{8cm}{\underline{\textbf{Bottleneck analysis}}: identifying and analyzing causes delaying progress (bottlenecks) on the student's educational trajectories.} & 
5 &
\citep{4_Mykola} \citep{2_Anuwatvisit} \citep{12_Caballero}\\ 
&&
\citep{17_Martinez} \citep{Potena23}
\\ \hline

\multirow{3}{8cm}{\underline{\textbf{Stopout/dropout analysis}}: identifying and analyzing causes that lead to stopout and dropout of students.} & 
8 &
\citep{4_Mykola} \citep{14_Salazar}\\ 
&&
\citep{15_Salazar} \citep{17_Martinez}\\
&&
\citep{19_Salazar} \citep{20_Salazar}
\\ 
&&
\citep{Diamantini24b} \citep{Diamantini24} 
\\ \hline

\multirow{3}{8cm}{\underline{\textbf{Trajectory/curriculum recommendation}}: making personalized recommendations to the student's educational trajectories or the curricular design to improve the expected results.} & 
8 &
\citep{4_Mykola} \citep{24_Awatef} \citep{11_Schulte}\\ 
&&
\citep{13_Wang} \citep{21_mahamed}\\
&&
\citep{22_Wagner} \citep{Rafied23} \citep{Roepke24}
\\ \hline
\end{tabular}
}
\label{tab:objectives}
\end{table}

Research objectives can take the form of questions. As analyzed in \citep{Zerbato22}, whether the PM methodology proposes to start with concrete questions to answer or if there is an exploratory analysis addressing general process-related and data-related questions, a project iteratively refines and generates domain-specific questions of interest. Some works \citep{11_Schulte,4_Mykola,7_Awatef,9_Cameranesi,8_Wang,20_Salazar,Potena23,Diamantini24} describe curricular analysis specific questions of interest, that can be organized into key themes as depicted in Table \ref{tab:mappingQuestions}.

\setlength\extrarowheight{2pt}
\begin{table}[!ht]
\caption{Research Questions and Related Process Mining Activities}
\centering
{\scriptsize
\begin{tabular}{|p{8.5cm}|p{8.5cm}|}
\hline 
\textbf{Question} & \textbf{Related Process Mining Activity} \\ \hline \hline

\rowcolor{lightgray}\multicolumn{2}{|l|}{\textbf{Curriculum Structure and Pathways}} \\ \hline
What is the real academic curriculum (study program)? & Discovery – Uncovering the pathways students take in the curriculum, capturing the real academic curriculum as implemented. \\ \hline
Is there a typical or optimal way to traverse the curriculum? & Discovery and Enhancement – Identifying common patterns (typical pathways) and suggesting optimal traversal sequences. \\ \hline
Do current prerequisites between courses make sense? & Conformance Checking – Verifying if prerequisites match actual patterns of student success and curriculum requirements. \\ \hline
Are specific curriculum constraints being followed? & Conformance Checking – Checking student adherence to curriculum constraints, such as prerequisites or course load limits. \\ \hline \hline

\rowcolor{lightgray}\multicolumn{2}{|l|}{\textbf{Student Behavior Patterns and Progression}} \\ \hline
Are there significant differences in behavior between students who graduated versus those who did not? & Discovery and Enhancement – Comparing trajectories of students to identify critical differences. \\ \hline
Do behavior patterns vary significantly between generations? & Discovery and Enhancement – Analyzing cohort-based differences to identify shifts or trends in behavior over time. \\ \hline
Are there identifiable best and worst practices among students regarding course-taking behavior? & Discovery and Prescriptive Analytics – Identify patterns and use this to guide students toward best practices. \\ \hline
Are students following the recommended study program more likely to graduate on time? & Conformance Checking and Predictive Analytics – Evaluating if adherence to the program correlates with timely graduation. \\ \hline \hline

\rowcolor{lightgray}\multicolumn{2}{|l|}{\textbf{Impact of Early Progress and Predictive Outcomes}} \\ \hline
How does students' progression in their first year impact final grades and time to graduation? & Predictive Analytics – Analyzing early progress as a predictor of long-term academic outcomes. \\ \hline
How likely is it that a student will successfully complete their studies or drop out? & Predictive Analytics – Estimating graduation or dropout likelihood based on students' early progress and trajectory patterns. \\ \hline
For students who drop out later, what are the typical trajectories, especially concerning high-failure courses that must be retaken? & Discovery and Enhancement – Mapping dropout patterns to identify common sequences and factors, like high-failure courses. \\ \hline
Are there early warning signs or specific progression patterns that indicate whether a student might need additional support? & Predictive Analytics – Detecting early signs in student trajectories that predict intervention needs. \\ \hline \hline

\rowcolor{lightgray}\multicolumn{2}{|l|}{\textbf{Bottlenecks and Performance Optimization}} \\ \hline
Are there patterns indicating bottlenecks in the curriculum? & Discovery and Enhancement – Locating frequent delays or bottlenecks where students struggle within the curriculum. \\ \hline
What are the most advantageous courses for students to take next, based on their progress? & Prescriptive Analytics – Recommending next courses for students based on historical patterns and successful pathways. \\ \hline
Are there specific course trajectories that can help students maximize their Grade Point Average (GPA)? & Prescriptive Analytics – Advising students on pathways that historically lead to high GPAs. \\ \hline \hline

\rowcolor{lightgray}\multicolumn{2}{|l|}{\textbf{Time to Completion and Graduation Outcomes}} \\ \hline
What is a student’s expected time to complete their degree? & Predictive Analytics – Estimating completion time based on individual student progress and historical data trends. \\ \hline
How does adherence to the program influence on-time graduation? & Conformance Checking and Predictive Analytics – Assessing if following the study program correlates with on-time graduation. \\ \hline
Can students expect similar outcomes the following semester, or should they adjust their workload to improve performance? & Predictive Analytics and Prescriptive Analytics – Forecasting next-semester performance and advising on workload adjustments. \\ \hline \hline

\rowcolor{lightgray}\multicolumn{2}{|l|}{\textbf{Decision-Making Support}} \\ \hline
Is it better to drop a course rather than predictably fail? & Predictive Analytics and Prescriptive Analytics – Using predictions to advise on dropping courses to maintain overall progress. \\ \hline
Should students adjust their workload to improve performance? & Predictive Analytics and Prescriptive Analytics – Recommending commitment adjustments based on expected performance. \\ \hline

\end{tabular}}
\label{tab:mappingQuestions}
\end{table}

Two key observations arise from these questions. First, we can distinguish between \textbf{student-related and curriculum-related questions}, each focused on different stakeholders and requiring insights at varying levels of detail. For instance, ``What is a student’s expected time to complete their degree?'' provides individualized insights for students about their likely time to completion, while ``How does adherence to the program influence on-time graduation?'' addresses aggregate-level insights aimed at understanding whether the study plan effectively supports timely graduation. This distinction reflects the need for \textbf{granularity of insights for stakeholders}, where some analyses are tailored to individual students while others serve curriculum designers and administrators. Second, while some questions align directly with specific objectives, such as using curricular conformance checking to assess whether curriculum constraints are followed, others require a synthesis of multiple analyses. For instance, identifying best and worst practices in student course-taking behavior would likely involve combining insights from trajectory discovery, bottleneck analysis, and success pattern identification. This need for \textbf{direct versus derived insights} highlights the importance of an integrated analytical approach that combines outputs from multiple techniques.

Additionally, these questions vary in terms of their \textbf{short-term versus long-term focus}, with some questions being short-term and action-oriented (e.g., ``Is it better to drop a course rather than predictably fail?'') and others focusing on long-term, structural analyses (e.g., ``Are specific curriculum constraints being followed?''). Segmenting questions by timeframe can help prioritize analyses based on urgency or impact. Similarly, \textbf{predictive questions} are aiming to forecast outcomes (e.g., ``How likely is it that a student will successfully complete their studies
or drop out?'') and \textbf{descriptive questions} explaining past behaviors (e.g., ``Are there significant differences in behavior between students who graduated versus those who did not?''). Balancing these descriptive, diagnostic, and predictive insights is essential to understand current and future behaviors comprehensively.

Furthermore, several questions imply a need for \textbf{intervention versus observation}, where some aim to generate actionable recommendations (e.g., ``Should students adjust their workload to improve performance?'') and others are purely observational, focusing on understanding current behaviors without prescribing changes (e.g., ``Do current prerequisites between courses make sense?''). This distinction helps clarify when insights should lead to specific recommendations.

Another theme is \textbf{cohort and generation analysis}, which is evident in questions that explore behavior changes across different generations (e.g., ``Do behavior patterns vary significantly between generations?''). This type of question provides insights into curriculum effectiveness over time, potentially identifying the impacts of policy changes or shifts in student demographics.

Finally, many questions reflect a goal of \textbf{outcome optimization}, focusing on enhancing student outcomes, whether it’s maximizing GPA, reducing time to degree, or minimizing dropout risk. This overarching objective unifies student-centered and curriculum-centered inquiries under the shared purpose of improving educational outcomes.

\vspace{-0.5cm}

\subsection*{(RQ2) How PM techniques are used?}

In general terms, as described in \citep{4_Mykola}, curriculum mining refines the general PM tasks within the educational context: building a curriculum model from the observed student's behavior (discovery), checking whether this behavior matches the expected curricular definitions (conformance checking), and projecting information extracted from the observed data onto the model to perform further analysis and recommend further actions (extension/enhancement). 
Beyond traditional PM tasks, curriculum mining often incorporates predictive analytics to anticipate future outcomes (such as predicting dropout risk or delays) and prescriptive analytics to recommend actions (like adjusting course load or selecting specific courses). These elements go beyond static model construction by enabling proactive interventions. Table \ref{tab:mappingQuestions} provides insights into the potential relations between PM tasks and the answer to curriculum mining questions.

Table \ref{tab:summaryWorks} highlighting various aspects concerning the process mining techniques actually addressed by the studies (if specific information is provided in the papers).
It includes the process mining type addressed (i.e., Discovery (D), Conformance (C), Enhancement (E), Prediction/Prescriptive Analytics (P)), and the perspective involved. 
In terms of event semantics, different studies represent academic processes by individual courses taken by students or milestones, such as the completion of semesters or years, to capture progression patterns. The model types vary among works, including Petri Nets, Directly Follows Graphs (DFG), Dependency Graphs (DG), and other notations. For conformance techniques, alignment-based methods are commonly used, though sequence matching and partial orders are also applied in some cases.

\begin{table}[!ht]
\caption{Process mining techniques.}
\centering
{\footnotesize
\begin{tabular}{|c|l|l|l|l|l|}
\hline
\textbf{Paper} & 
\textbf{PM Type} &
\textbf{Perspective} & 
\textbf{Event Semantics} & 
\textbf{Model Type} & 
\textbf{Conformance} \\
\hline\hline

\citep{1_Trcka} &
D, C &
control-flow & 
courses & 
Colored Petri Net & 
Alignment-based 
\\\hline

\citep{2_Anuwatvisit} &
D, C, E &
control-flow, time & 
courses & 
Petri Net & 
Alignment-based 
\\\hline

\citep{3_Ayutaya} &
D &
control-flow & 
courses & 
Heuristics Net & 
- 
\\\hline

\citep{4_Mykola} &
D, C, E &
control-flow, time & 
courses & 
Colored Petri Net & 
Alignment-based 
\\\hline

\citep{24_Awatef} &
D, E &
control-flow, org. & 
courses & 
Social Network & 
- 
\\\hline

\citep{7_Awatef} &
D, C &
control-flow & 
courses & 
Heuristic Net & 
LTL-based 
\\\hline

\citep{25_Bendatu_Yahya_2015} &
C &
control-flow & 
milestones & 
- & 
Sequence matching 
\\\hline

\citep{8_Wang} &
D, C &
control-flow & 
courses & 
Colored Petri Net & 
-
\\\hline

\citep{9_Cameranesi} &
D, C &
control-flow & 
courses & 
Petri Net & 
Alignment-based
\\\hline

\citep{11_Schulte} &
D, P &
control-flow & 
courses & 
- & 
- 
\\\hline

\citep{12_Caballero} &
D, E &
control-flow, time & 
courses & 
- & 
- 
\\\hline

\citep{13_Wang} &
D, P &
control-flow & 
courses &
DG &
- 
\\\hline

\citep{14_Salazar} &
D &
control-flow & 
courses & 
DFG & 
-
\\\hline

\citep{15_Salazar} &
D &
control-flow & 
milestones & 
DFG & 
- 
\\\hline

\citep{17_Martinez} &
D, E &
control-flow, time & 
courses & 
Petri Net, DG & 
- 
\\\hline

\citep{19_Salazar} &
D &
control-flow & 
milestones & 
DFG & 
- 
\\\hline

\citep{20_Salazar} &
D &
control-flow & 
milestones & 
DFG & 
- 
\\\hline

\citep{23_Hobeck} &
D, C &
control-flow & 
milestones & 
DFG, Petri Net, BPMN & 
Alignment-based 
\\ \hline

\citep{21_mahamed} &
D, P &
control-flow & 
courses & 
- & 
- 
\\\hline

\citep{22_Wagner} &
D, C, P &
control-flow & 
courses & 
DFG, Petri Net, BPMN & 
Alignment-based 
\\\hline

\citep{Potena23} &
D, C, P &
control-flow & 
milestones, exams & 
Petri Net & 
Alignment-based
\\ \hline

\citep{Rafied23} &
D, P &
control-flow & 
courses & 
DFG, Decision Tree & 
-
\\ \hline

\citep{Diamantini24b} &
D, C, P &
control-flow & 
courses, milestones & 
Petri Net & 
Cost-based fitness 
\\ \hline

\citep{Diamantini24} &
D, C, E &
control-flow, time & 
courses, milestones & 
Causal Net & 
Alignment-based
\\ \hline

\citep{Puttow24} &
E &
control-flow &
courses & 
DFG & 
-
\\ \hline

\citep{Rennert24} &
D, C &
control-flow & 
courses, milestones & 
Workflow Net & 
Partial order alignments 
\\ \hline

\citep{Roepke24} &
D, C, P &
control-flow & 
courses & 
- & 
-
\\ \hline

\end{tabular}
}
\label{tab:summaryWorks}
\end{table}

\begin{figure}[!ht]\captionsetup[subfigure]{font=footnotesize}
\centering
\begin{subfigure}{.8\textwidth}
    \centering
    \includegraphics[width=\linewidth]{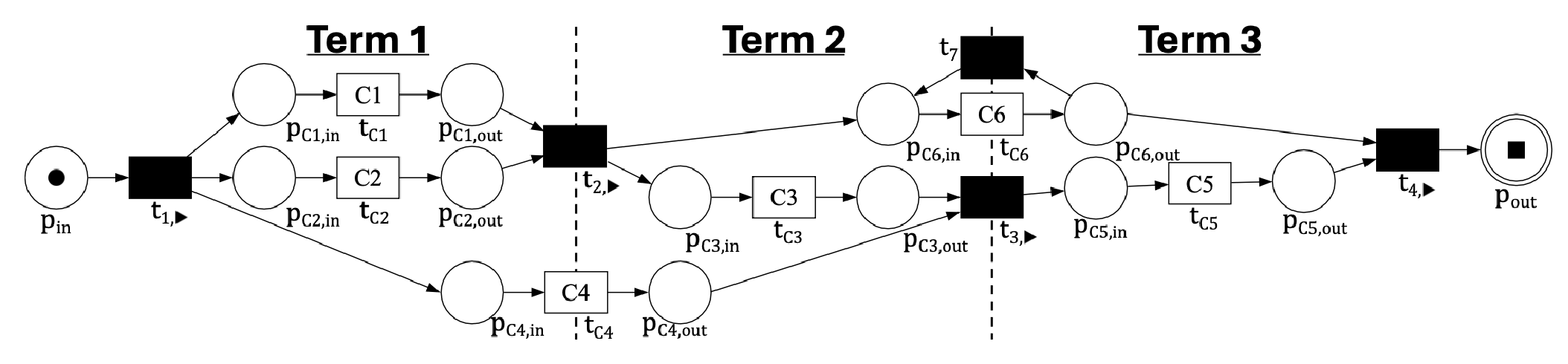}
    \caption{Worflow Net from \citep{Rennert24}  (using Pm4Py)}
\end{subfigure}
\vfill
\vspace{0.2cm}
\begin{subfigure}{.45\textwidth}
    \centering  
    \includegraphics[width=\linewidth]{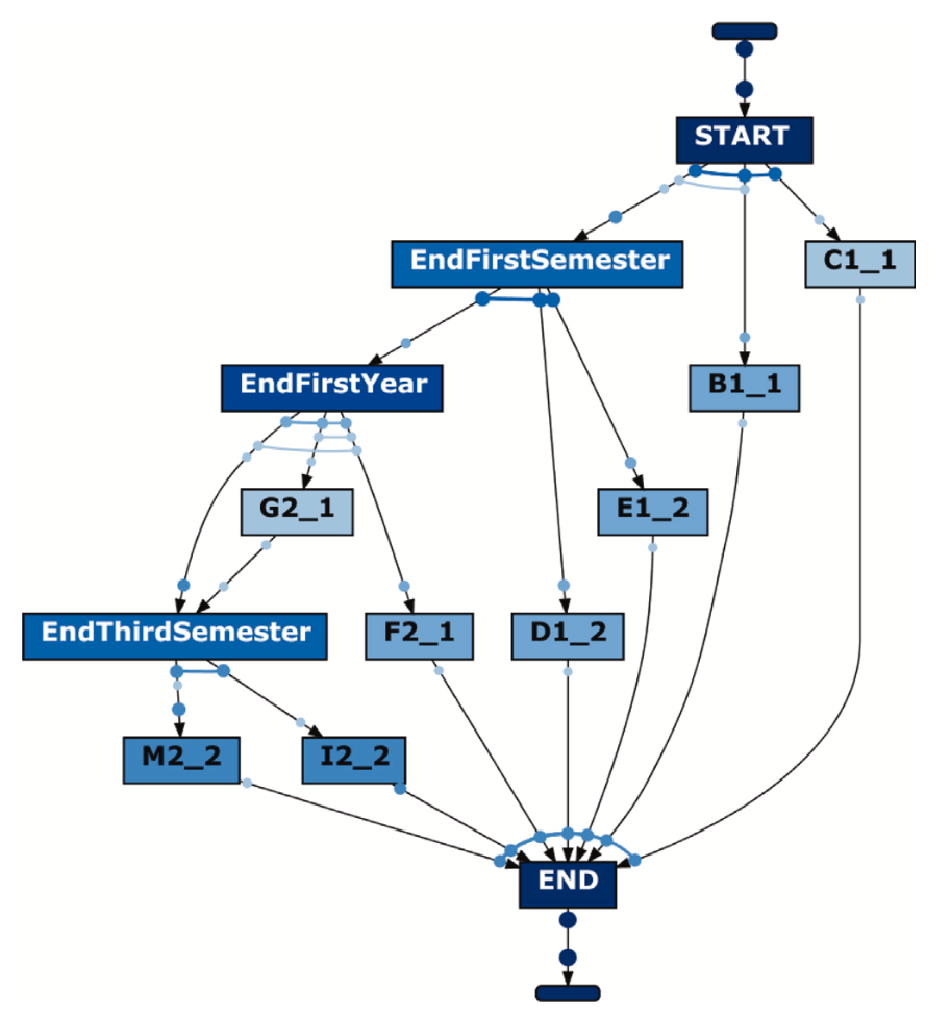}
    \caption{Causal Net from \citep{Diamantini24} (using ProM)}
\end{subfigure}
\begin{subfigure}{.4\textwidth}
    \centering  
    \includegraphics[width=0.6\linewidth]{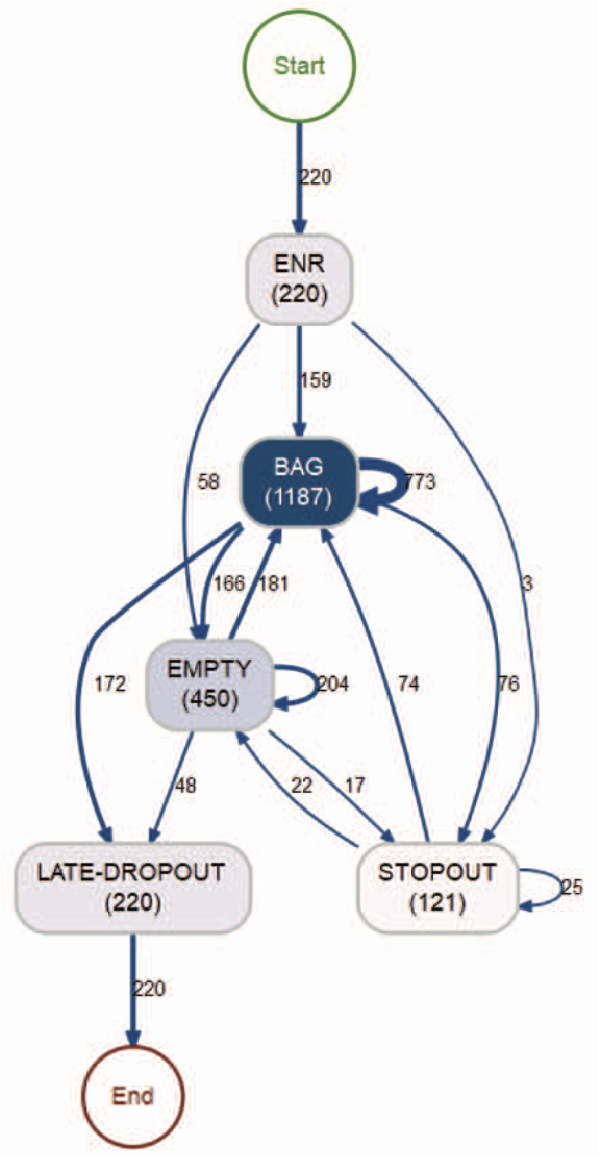}
    \caption{DFG from \citep{15_Salazar}  (using bupaR)}
\end{subfigure}
\caption{Example of process models for curricular analysis using different notations and event semantics.}
\label{fig:modelsexamples}
\end{figure}

As already described and summarized in Table \ref{tab:objectives}, the main objective addressed in the papers is the discovery of educational trajectories. It requires defining a process model that reflects the trajectory of students. Cases are usually actual students' trajectories in which events of the discovered process represent courses, e.g., in \citep{8_Wang,2_Anuwatvisit,3_Ayutaya,9_Cameranesi,11_Schulte,12_Caballero,17_Martinez,23_Hobeck}, or certain curriculum milestones such as the end of a semester \citep{Potena23,Diamantini24b,Diamantini24}. However, the semantics of an event can be conceived from different perspectives, not only from a curricular standpoint, e.g., the courses a student approves, but also from a contextual perspective, e.g., graduation in time, late dropouts or the number of pending studies. This interpretation is used in works focusing on stopout/dropout analysis, e.g., \citep{14_Salazar}, as we will describe later.
Figure \ref{fig:modelsexamples} depicts three different model notations with their own event semantics. The decision depends on different aspects. In \citep{Rennert24}, a curricular model represents courses (as transitions) with term start and end silent transitions, allowing the analysis of the temporal diﬀerence between the expected and actual course-taking term. In such a case, Workflow Nets are used since they allow the definition of a partial-order alignment algorithm using such notation formally. In \citep{Diamantini24}, there is a representation of courses and milestones (end of semester and end of a year). However, in such a case, the authors chose Causal Nets since it provides a more compact and simpler representation of the routing logic inferred from the discovery algorithm (i.e., the Heuristic Miner). Finally, in \citep{15_Salazar}, the authors model the context of a student as activities: enrollment, stopout/dropout, bag (i.e., pending failed courses), and empty (no pending courses). They use DFG since its notation is available in the statistical analysis environment they used.

Since a trajectory refers both to a student's path and a set of students, in some cases, there is an a priori clustering of traces based on considering additional information, e.g., demographic information or students' performance characteristics, which sometimes requires additional data mining tools \citep{14_Salazar,Diamantini24b}. In other cases, clustering is performed beforehand during the analysis. In \citep{24_Awatef}, the authors propose a clustering method, based on sequence clustering, for partitioning training paths based on key performance indicators (KPIs). Moreover, in \citep{9_Cameranesi}, the authors segmented traces into three disjoint groups depending on the final grade mark and length of the study period, using trace clustering. In both cases, partitioning first requires analyzing KPI and path lengths.

When logs contain low-level information unsuitable for analyzing the problem, e.g., to analyze how knowledge areas spread over several courses are covered, it is possible to associate more precise semantics to those events (semantic lifting), e.g., associating the knowledge areas to the events \citep{vanZelst2020EventAI}. In \citep{7_Awatef}, the authors propose a (semi)automatic procedure to associate semantics to training labels based on an ontology and then discover simplified training paths using the semantic information. This semantic lifting is also used in other works, such as \citep{14_Salazar}, not to discover a trajectory but to perform a stopout/dropout analysis from another perspective.

A curriculum process model can also be formally modeled from scratch. In \citep{1_Trcka}, the authors propose a method based on Colored Petri Nets patterns and show how pattern mining can be used to discover a curriculum model.

Curricular conformance conformance checking. As described in \citep{22_Wagner}, conformance is used in two senses: first, to identify how different cohorts behave and to identify outliers, and second, to measure adherence to a planned curriculum. As described in \citep{23_Hobeck}, trajectories have a high variance. Thus, two strategies are followed: clustering of traces once a cohort or specific group of students is identified, as described before, and projecting into the model and the log partial information or stages in the curricula, e.g., compulsory or first-year courses. 
The main conformance-checking technique used is alignment. However, there are a couple of exceptions. In \citep{7_Awatef}, the authors propose using LTL expressions to check whether trajectories conform to specific requirements such as minimal paths and business rules. In \citep{25_Bendatu_Yahya_2015}, the authors check the curriculum design using sequence matching analysis to the actual trajectories using sequence matching analysis. Finally, in \citep{Diamantini24b}, the authors use a cost-based conformance-checking technique leveraging the fitness metric by considering the cost of skipping and inserting activities. An additional perspective to conformance-checking is followed in \citep{Rennert24} in which partial orders are used to relax the constraints of strictly ordered traces, and alignments can be computed without requiring a strict order in which courses are oﬀered.

In the case of bottleneck analysis, in all works \citep{2_Anuwatvisit,12_Caballero,17_Martinez,Potena23,Diamantini24,Diamantini24b}, a process model directly represents the students' actual trajectories, i.e., each event is a course or exam taken or approved. Since time matters, it is firmly based on performance analysis. In \citep{2_Anuwatvisit,17_Martinez}, the authors identify courses that prevent academic progress by performing a waiting time analysis and identifying courses that exhibit approval times above the expected. 
In \citep{12_Caballero}, the authors obtained frequencies of students according to the time students invest in completing a course and extracted requirements that define what a bottleneck is from the observed behavior, e.g., courses with frequencies for total passed students lower than 50\%. In \citep{Potena23,Diamantini24,Diamantini24b}, the authors detect exam-taking patterns and use prediction models to elucidate the impact of different factors on graduation times.

In the case of stopout/dropout analysis, a process model can directly represent the students' courses taken or approved. In \citep{17_Martinez}, the authors follow this idea and identify the last courses with which students graduate or drop out. They consider if two years have passed since the last approval, a rule defined by their university. However, other works use different semantics for events. In \citep{14_Salazar}, the authors built an event log at the semester level, converting approval data to higher-level events such as a \textit{STOPOUT} event if the student did not enroll in courses in the semester and a \textit{BAG} event if the student enrolls but has pending courses, among others. They also consider final events such as \textit{EARLY-DROPOUT}, for students who dropped out no later than the fourth semester, and \textit{LATE-GRAD} for students who took longer to graduate than the time defined for the program, among other events. In \citep{15_Salazar}, the authors add other events of interest to see how pending courses accumulate in a semester and if that affects the stopout in the long run. They also studied how gender, income, and entry math skills may explain differences in trajectories. In \citep{19_Salazar}, they refine this idea by defining a model according to the courses students have failed and have yet to pass. Finally, they focus on dropout in \citep{20_Salazar}, using a state diagram that shows the possible result events in the retake-stopout model.

Enhancement techniques usually overlay additional information in a process model to enhance comprehension of aspects. In \citep{Puttow24}, the authors introduce a method to associate different context representations of a data record to a label, i.e., the coloring of activities based on the case or event attributes and activity frequency. It differentiates many possible representations in the same data instance, e.g., color ranges represent the frequency of each course taken or the proximity to the minimum and maximum grades.

Finally, trajectory/curriculum recommendations, such as in \citep{4_Mykola,11_Schulte,22_Wagner,Roepke24}, are based on combining several strategies. In general terms, as described in \citep{22_Wagner,Roepke24}, PM techniques are used to identify successful students' paths and to detect deviations from expected plans. This information is combined with rule-based AI strategies to derive rules as recommendations of more suitable study paths with higher success rates, i.e., predictions with a certain confidence level. Many possible strategies are applied, e.g., in \citep{13_Wang}, the authors apply sequence-based approaches (e.g., sequential pattern mining); in \citep{21_mahamed}, the authors use logistic regression for resolving a classification problem; and in \citep{Rafied23}, the authors derive decision tree models.
In a less restrictive curricular context, one possibility is to discover the trajectories that students naturally follow and define a curriculum based on that. In \citep{24_Awatef}, the authors analyze social networks between courses and training providers of a consulting company, e.g., to discover popular providers. They also propose a clustering method for partitioning training paths based on key performance indicators and use discovery to analyze such clusters.

\vspace{-0.5cm}

\subsection*{(RQ3) What research methodologies and PM methods and tools are used?}

Table \ref{tab:summaryTools} provides a comparative view of process mining methods, tools, and algorithms used across the primary studies. 
Each entry includes the study's title and the information explicitly provided concerning the methodology followed for the case study analysis (if applicable), the tool the authors developed (if applicable), and the base tools and discovery algorithms used.

\begin{table}[!ht]
\caption{Process mining methods, algorithms, and tools.}
\centering
{\footnotesize
\begin{tabular}{|c|l|l|l|l|}
\hline
\textbf{Paper} & 
\textbf{Method} &
\textbf{Tool} & 
\textbf{Base tool} & 
\textbf{Algorithm} \\ 
\hline\hline

\citep{1_Trcka} &
--- &  
StudyBuddy / BuddyAnalytics &
ProM, CPN Tools &
---
\\\hline

\citep{2_Anuwatvisit} &
Ad hoc &
--- &
ProM &
---
\\\hline

\citep{3_Ayutaya} &
Ad hoc &
--- &
ProM &
Heuristic Miner
\\\hline

\citep{4_Mykola} &
--- &
CurriM &
ProM, CPN Tools &
---
\\\hline

\citep{24_Awatef} &
--- &
PHIDIAS &
ProM &
Fuzzy Miner, Social mining
\\\hline

\citep{7_Awatef} &
--- &
--- &
ProM, WSML 2 Reasoner, LTL
checker &
Heuristic Miner
\\\hline

\citep{25_Bendatu_Yahya_2015} &
Ad hoc &
Unnamed &
ProM &
---
\\\hline

\citep{8_Wang} &
--- &
--- &
--- &
---
\\\hline

\citep{9_Cameranesi} &
--- &
--- &
ProM &
Infrequent Inductive Miner
\\\hline

\citep{11_Schulte} &
--- &
Unnamed &
ProM, Disco, KNIME, R, Shiny &
---
\\\hline

\citep{12_Caballero} &
Ad hoc &
--- &
ProM &
---
\\\hline

\citep{13_Wang} &
--- &
Unnamed &
ProM &
---
\\\hline

\citep{14_Salazar} &
PM\textsuperscript{2} &
--- &
bupaR, R &
---
\\\hline

\citep{15_Salazar} &
PM\textsuperscript{2} &
--- &
bupaR, R &
---
\\\hline

\citep{17_Martinez} &
--- &
--- &
ProM &
Heuristic Miner, Inductive Miner
\\\hline

\citep{19_Salazar} &
PM\textsuperscript{2} &
--- &
bupaR, R &
---
\\\hline

\citep{20_Salazar} &
PM\textsuperscript{2} &
--- &
bupaR, R &
---
\\\hline

\citep{23_Hobeck} &
PM\textsuperscript{2} &
--- &
ProM, Disco &
---
\\ \hline

\citep{21_mahamed} &
--- &
Unnamed &
--- &
---
\\\hline

\citep{22_Wagner} &
--- &
StudyBuddy / BuddyAnalytics &
--- &
---
\\\hline

\citep{Potena23} &
--- &
--- &
Pm4Py &
Infrequent Inductive Miner
\\ \hline

\citep{Rafied23} &
--- &
AIStudyBuddy &
Pm4Py &
---
\\ \hline

\citep{Diamantini24b} &
--- &
--- &
Pm4Py &
Infrequent Inductive Miner
\\ \hline

\citep{Diamantini24} &
--- &
--- &
ProM &
Heuristic Miner
\\ \hline

\citep{Puttow24} &
--- &
--- &
--- &
---
\\ \hline

\citep{Rennert24} &
--- &
StudyBuddy / BuddyAnalytics &
Pm4Py &
---
\\ \hline

\citep{Roepke24} &
--- &
StudyBuddy / BuddyAnalytics &
--- &
---
\\ \hline

\end{tabular}
}
\label{tab:summaryTools}
\end{table}

As tangible results of the works, most of them provide real-world application examples by taking historical information from universities and consulting companies supporting software systems. From a methodological perspective, since 2019, some works have started using the PM\textsuperscript{2} \citep{PM2} methodology to guide the execution of their PM projects; before that, only a few works described using an ad hoc methodology.
Although a methodology exists, the case studies do not always follow a systematic approach, and the methodology is not tailored for application in the specific context.

Only some works define comprehensive tools supporting curricular mining initiatives. In \citep{4_Mykola}, the authors envision CurriM, a tool covering many PM actions about curricular models. Moreover, in \citep{22_Wagner,Roepke24}, the authors describe an approach and tool to support individualized study planning and monitoring using campus management system data and study program models. In this case, the StudyBuddy tool provides a planning tool for students, while the BuddyAnalytics offers an analytics dashboard for program designers. Unfortunately, none of these tools are available. However, the project described in \citep{22_Wagner,Roepke24} is ongoing. Other works also refer to unavailable developed tools supporting their specific analysis.

Before 2017, the most used tool for performing analysis was ProM. Some works describe the discovery algorithms used, i.e., Fuzzy Miner, \citep{7_Awatef}, Heuristic Miner \citep{7_Awatef,17_Martinez} and infrequent Inductive Miner (iIM) \citep{9_Cameranesi}. Moreover, in \citep{3_Ayutaya}, the authors analyze the appropriateness of the Heuristic Miner for process discovery. No other work compares algorithms to define the most suitable one in this context. The decision to use one or the other is usually not clearly justified, except in cases where the use of a specific notation is defined that favors some aspect, e.g., visualization of the results \citep{Diamantini24}, or formal bases that allow defining additional algorithms \citep{Rennert24}.
Since 2017, there has been a slight change in the existence of other PM tools, i.e., Disco and bupaR. Also, in \citep{11_Schulte}, the authors describe the combination of PM and DM tools they used: KNIME, Disco, ProM, and R. Finally, in \citep{8_Wang}, the authors developed a curriculum simulator to mimic the behaviors of different kinds of students, which is unavailable.

\section{Challenges \& opportunities}\label{sec:discussion}

In this section, we discuss \textbf{(RQ4) What are the open challenges and research opportunities?} There are general PM challenges from a technical \citep{ProcessMiningBook} and organizational \citep{MartinFKGLRADRW21} perspective, as well as more specific EPM challenges \citep{SLR2} that must be addressed. In what follows, we summarize and complement specific curricular analysis challenges described in primary studies.

\begin{description}
    \item[Log abstraction] Log abstraction is a core preprocessing task in PM that simplifies logs by aggregating events, defining event classes, and clustering \citep{Liu24}. In curricular analysis, however, process models must represent student trajectories at different levels of granularity, such as course-level activities versus curriculum milestones. An open challenge is to evaluate whether existing log abstraction techniques sufficiently capture these diverse granularity levels or if tailored, domain-specific methods are needed, as suggested by \citep{Puttow24}. Ensuring granularity and interpretability is critical to understanding student behavior across different levels.

    \item[Heterogeneity and complexity.] PM tools must handle voluminous data from educational information systems. This data describes heterogeneous and complex situations since educational processes reflect the high diversity of behaviors in students' learning paths \citep{24_Awatef}. Additionally, some undesired behaviors from the PM perspective are not extraordinary, e.g., non-finalized cases of dropout students. Moreover, as described in \citep{23_Hobeck}, curricula are often flexible, particularly when there is a wide variety of compulsory courses and when exceptions are considered, e.g., exams could be repeated after failing, or there are lateral program entries when a student switches its career or spend a semester studying in a foreign program. All these aspects affect process discovery (generating spaghetti models) and conformance checking. 
    Addressing this challenge requires strategies such as clustering student cohorts with similar characteristics, using selective filtering to simplify process models without losing critical information, and developing more tailored conformance-checking techniques.
    
    Cluster analysis aims to identify typical behavior patterns of groups of similar students based on intra- and inter-course data such as examination results and order of the courses in their trajectories \citep{li2019educational,krivzanic2020educational}. 
    In \citep{7_Awatef,9_Cameranesi}, the authors mentioned the need to develop new and refined clustering and classification techniques and to compare results from different techniques in the context of curricular analysis.
    Clustering and filtering strategies could be applied on demand within an iterative analysis, e.g., performing trace clustering on students with similar characteristics and filtering a subset of such traces by looking at other specific student attributes.
    Another strategy is aggregating events when the raw data might need to be more cohesive or only consider partial information or stages in the curricula, e.g., compulsory or first-year courses, when performing the analysis. As noticed in \citep{23_Hobeck}, the use of filters to reduce the noise of the log and the resulting models comes at the cost of lowering their expressiveness. 

    New conformance-checking techniques could be analyzed, e.g., considering partial orders when computing alignments to make results less prone to the order in which courses are oﬀered \citep{Rennert24}

    \item[Integrated analysis strategies.] Some analyses, e.g., analyzing dropout, sometimes require links with other contextual aspects, such as demographic information of students and the context in which a course is offered (e.g., the number of teachers assigned to a course). Two challenging requirements arise: enriching logs with additional data and incorporating multidimensional and integrated analysis strategies between PM and data mining (DM), as devised in \citep{DCM0T21}. 
    DM allows projecting information from the observed data onto the models to understand underlying relationships \citep{4_Mykola}. In this sense, it is possible not only to project simple demographic factors \citep{16_Goel} such as gender, age, nationality, ethnicity, family education, or economic position but to complement and complete curricula analysis using the educational history of the students and vocational ambitions to explain differences in trajectories \citep{sakurai2012case}, and econometric models to quantify the influence of variables along time \citep{14_Salazar}. DM can also enrich the analysis of university curricula and the acquired competencies, for example, adding Tracer Study (TS) data of their graduates, such as the waiting period for getting the first job, the success of graduates competing in the selection, earned salaries, and graduates' opinions \citep{arifin2022using}. Process Cubes \citep{BoltA15} and object-centric process mining \citep{Aalst19} could be explored in this multidimensional context for data management and log generation. 
    Consider adding an analysis of external factors that may affect students' educational trajectories could help distinguish between curricular and non-curricular reasons behind delays or dropouts.

    \item[Curricular flexibility and evolution.] Student diversity in higher education and the need to deliver students who adapt to changes has made educational institutions address these needs by designing a flexible curriculum \citep{jonker2020curriculum}. To answer if current prerequisites make sense, it is necessary to evaluate what would happen if these prerequisites were not present, something that can only be approximated based on assumptions or through the generation of simulations. Recent studies based on DM and simulation present generalized approaches to exploring the impact of policy changes on the curriculum and creating effective and individually verifiable simulations without requiring a rigid curriculum graph to evaluate changes to flexible curricula \citep{baucks2022simulating}. Process simulation could be applied to possible trajectories based on the history of discovered trajectories. It would also be interesting to compare the results of both types of simulation to assess their potential. 

    Another related concept in the context of curricular evolution is concept drift \citep{24_Awatef,4_Mykola,11_Schulte}, i.e., a significant change in a process over time. PM is particularly sensitive to this aspect, and more profound studies still need to be developed in the educational context.    

    \item[Prediction.] DM techniques have been widely used to predict students' performance and dropout, using demographic and educational attributes, applying techniques such as Decision Trees, Random Forest, Bayes Network, and others \citep{alshareef2020educational}. 
    Integrating these predictive models with PM techniques is essential to leverage the strengths of both methods and improve prediction accuracy. Comparative analysis of PM and DM results is also needed to determine best practices for combining methods in educational prediction contexts.

    \item[Metrics.] Using metrics provides objective information on which to evaluate the current state of a curriculum and base decisions regarding policy changes. It could be possible to combine the discovered trajectories with specific curriculum metrics \citep{Ochoa2016SimpleMF}. Only a few works, such as \citep{22_Wagner}, have used Key Performance Indicators (KPIs) about students, study programs, or cohorts, such as the success rate of a course and exams passed in a semester. Further works could explore and evaluate the use of such information and determine its potential for integrated analysis.

    \item[Methodologies.] At a methodological level, some works use a well-defined but general methodology, e.g., PM\textsuperscript{2}, for guiding a curricular analysis project. We may extend the PM\textsuperscript{2} process to conduct concrete curricular analysis activities. It would imply defining specific actions and recommending particular techniques to be applied in each step, according to predefined questions of interest for curricular analysis. Moreover, we can consider a methodological context with an analysis based on existing combined DM and PM methods, as in \citep{DCM0T21}. Curricular analysis also requires an interdisciplinary approach that merges education, psychology, and data science insights, enabling more relevant and actionable results.

    \item[Tools.] Tools must be available if the curricular analysis is expected to be interpreted by non-technical users, e.g., CurriM \citep{4_Mykola} and StudyBuddy and BuddyAnalytics \citep{22_Wagner,Roepke24}. Many opportunities exist, e.g., providing real-time analysis tools and extending an existing PM tool with a specific layer of curricular analysis techniques guided by a concrete methodology.     
    Providing adequate visualization techniques and notations for improving the interpretation of results by users is also a challenge, as identified in \citep{24_Awatef,SLR2,23_Hobeck}.
    There is also room for improvement in applying predictive PM to provide recommendations. Students could be able to simulate an ideal trajectory and predict the probability of success based on it.
    
    \item[Practical application.] Several aspects make replication of studies hard, e.g., no standard methodology is followed for the same guiding questions, there are many alternatives to perform the analysis, data mining and PM strategies are usually disconnected, there is heterogeneous data, and no standard log format/content is defined, e.g., which demographic attributes should be considered for a given study. Some works, such as \citep{22_Wagner}, describe more standardized aspects, e.g., a general data model for PM initiatives independent of academic systems and programs. Given that several referred study programs have similarities, based on a standard methodology and a set of predetermined studies, it would be interesting to perform a cross-university curricular analysis to detect similarities and differences. 

    Stakeholder feedback and validation are vital to ensuring the practical application of findings. This objective includes developing a feedback mechanism for sharing insights with curriculum designers, faculty, and advisors. These people can use the results to make informed decisions and validate the effectiveness of changes based on trajectory and curriculum recommendations.
    
    Finally, all these analyses must be conducted strongly emphasizing ethical and privacy considerations. Ensuring data privacy and ethical use, particularly with sensitive student data, is critical. This objective outlines the methods for de-identifying data and securing permissions, supporting rigorous and ethical analysis.
\end{description}

Based on the current state of research and the open challenges we have described, we can identify several specific research opportunities for curricular mining summarized in Table \ref{tab:summaryResearchOpportunities}. The main research directions include developing advanced data preprocessing methods tailored to educational data and addressing ethical considerations. Other directions focus on creating flexible, scalable, and interpretable models, integrating PM with predictive analytics to support at-risk students, and refining visualization tools in this specific domain (although much progress has been made in general-purpose mining tools). Developing adaptive recommendation systems and simulation techniques can also help model curricular changes. Cross-institutional analysis, metric development, stakeholder feedback, and concept drift analysis are essential for standardized methodologies and long-term adaptability, enabling more effective curriculum management and student support.

\begin{table}
\caption{Summary of research opportunities.}
\centering
{\footnotesize
\begin{tabular}
{|p{0.2\textwidth}|p{0.5\textwidth}|p{0.2\textwidth}|}
\hline
\textbf{Category} &
\textbf{Research Opportunity} & \textbf{Related Challenges} \\
\hline\hline

\multirow{2}{0.2\textwidth}{\textbf{Data Preprocessing and Semantic Integration}} 
& Develop tailored abstraction methods for curricular analysis. & \multirow{3}{0.2\textwidth}{Heterogeneity and Complexity, Log abstraction} \\
& Integrate semantic context with models to improve interpretability. & \\
& Design clustering methods to group student patterns and focus on specific subgroups. & \\ 
\hline

\multirow{2}{0.2\textwidth}{\textbf{Addressing Ethics and Privacy Concerns}} 
& Develop specific de-identification techniques and data security frameworks to ensure privacy compliance. & \multirow{2}{0.2\textwidth}{Practical Application} \\
& Establish ethical guidelines for transparent and responsible use of data. & \\
\hline

\multirow{2}{0.2\textwidth}{\textbf{Flexible and Scalable Modeling }} 
& Design hybrid models that balance flexibility and simplicity. & \multirow{2}{0.2\textwidth}{Curricular Flexibility, Heterogeneity and Complexity} \\
& Develop adaptive models that automatically adjust to diverse educational contexts. & \\
\hline

\multirow{2}{0.2\textwidth}{\textbf{Predictive Modeling and Early
Intervention}}
& Develop predictive models for dropout risks and academic support. & \multirow{3}{0.2\textwidth}{Prediction} \\
& Incorporate detailed student data to improve accuracy in early warning systems. & \\
& Create early warning systems for timely dropout prevention. & \\
\hline

\multirow{2}{0.2\textwidth}{\textbf{Visualization and Bottleneck
Analysis Tools}} 
& Build real-time visualization tools to identify bottlenecks with interpretative insights. & \multirow{2}{0.2\textwidth}{Heterogeneity and Complexity} \\
& Use ML to predict bottlenecks based on historical data. & \\
\hline

\multirow{2}{0.2\textwidth}{\textbf{Conformance Checking and
Simulation for Flexible Pathways}} 
& Develop metrics for varied student paths, focusing on sequence and performance. & \multirow{3}{0.2\textwidth}{Curricular Flexibility and Evolution} \\
& Research hybrid approaches to balance adherence and outcomes. & \\
& Use simulations to model potential impacts of curriculum changes on pathways. & \\
\hline

\multirow{4}{0.2\textwidth}{\textbf{Standardized Methodologies,
Tools and Stakeholders
Feedback}}
& Create tools with visual and predictive capabilities accessible to non-technical users. & \multirow{6}{0.2\textwidth}{Integrated Analysis Strategies, Methodologies, Practical Application, Tools} \\
& Develop interactive dashboards for student path exploration. & \\
& Include feedback systems for curriculum designers, faculty, and advisors. &\\
& Adapt PM\textsuperscript{2} methodology to guide educational analysis. & \\
& Develop validation methods and iterative feedback loops to ensure practical use of PM recommendations. & \\
& Explore interdisciplinary methodologies to enhance results. & \\
\hline

\multirow{2}{0.2\textwidth}{\textbf{Adaptive and Context-Aware Recommendations}} 
& Research personalized recommendations based on student progression. & \multirow{3}{0.2\textwidth}{Heterogeneity and Complexity, Integrated Analysis Strategies} \\
& Include demographic and external factors like financial or employment status to distinguish curricular impacts. & \\
&
Develop multidimensional analysis approaches, e.g., exploring process cubes and object-centric process mining. &\\
\hline

\multirow{2}{0.2\textwidth}{\textbf{Metric Development and Success Evaluation}} 
& Define outcome-based metrics (e.g., graduation rates) for curriculum effectiveness. & \multirow{2}{0.2\textwidth}{Metrics} \\
& Measure the impact of curriculum changes on success metrics. & \\
\hline

\multirow{2}{0.2\textwidth}{\textbf{Concept Drift Analysis}} 
& Create methods to detect changes in curricular processes over time. & \multirow{2}{0.2\textwidth}{Curricular Flexibility and Evolution} \\
& Develop adaptive models that respond to ongoing curriculum updates. & \\
\hline

\multirow{2}{0.2\textwidth}{\textbf{Cross-Institutional interdisciplinary Analyses}}
& Standardize data models and methodologies to enable cross-university comparisons. & \multirow{2}{0.2\textwidth}{Heterogeneity and Complexity, Methodologies, Practical Application}  \\
& Develop a toolbox with basic techniques and guidelines for studying the main problems associated with curriculum analysis. & \\
\hline

\end{tabular}
}
\label{tab:summaryResearchOpportunities}
\end{table}
\section{Threats to validity}\label{sec:threats}

This section presents the threats to validity \cite{Wohl12}: construction, internal, conclusion, and external.

As a threat to construct validity, the selection of sources or the search string may not retrieve all relevant articles. To minimize them, we selected recognized sources based on their coverage of international publications from conferences and top-level journals in the area, and we opted for a search string that maximized the inclusion of articles using ``process mining and education*'' as critical terms. Exploratory searches were carried out in the sources and discussion among the authors for prior adjustments.

Reproducibility and bias in the selection of studies threaten internal validity. To minimize these, the protocol was specified, including the sources, the search string, and the inclusion/exclusion criteria, and the results were discussed between the authors at each stage to prevent possible selection bias.

As a threat to external validity, the quantity and relevance of the selected studies can affect the generalization of results. To minimize this threat, the articles were selected using the inclusion/exclusion criteria defined in the protocol and discussions between the authors. Although the number of chosen articles compared to those recovered may seem low, the review's focus is specifically on EPM over curricula analysis, which narrows the area. The results presented represent the topic's current state to our knowledge. There could be articles that need to be included that were not published or indexed at the time of the searches.

Dependence on the authors' perspective of the review threatens the validity of the conclusions. To minimize this threat, the information recovered from the analyzed articles is the one indicated in each proposal. If different visions were used to extract the identified categories, these were discussed between the authors.
\section{Conclusions}\label{sec:conclusions}

We have summarized the results of a systematic literature review that identifies works on applying PM to curricular analysis. It has been a small but active research area in the last ten years, encompassing the PM community's development.
The reviewed studies encompass the main types of PM—discovery, conformance checking, enhancement, and predictive/prescriptive analysis—and have been organized into five key categories: from uncovering educational trajectories to identifying issues such as dropout and stopout and generating evidence-based recommendations.

In general, the objectives cannot be reached using PM in isolation. Many other data mining and AI techniques could be used in combination, such as clustering of traces for different student groups to perform more specific analysis, the definition of KPIs to express target educational goals, and using AI to derive
rules as recommendations for higher success rate study paths.

Despite progress, several research opportunities remain. Strengthening connections with Educational Data Mining (EDM) could enhance the depth of curricular analysis. Given the sensitive nature of student data, ensuring data privacy and addressing ethical concerns are critical considerations. Developing frameworks for secure data management and ethical guidelines would support the responsible use of information while respecting student privacy. Improvements in methodological frameworks are also essential to structure curricular analyses around targeted research questions. Flexible and scalable PM models that can adapt to the diversity and complexity of educational trajectories are essential to accommodate various institutional contexts and student behaviors. Another promising area is predictive modeling for early intervention, enabling institutions to identify at-risk students and take timely action based on early dropout or academic struggle indicators.

Additionally, developing openly accessible tools for non-technical users could lower implementation barriers, making PM-based insights more widely applicable within educational institutions. Real-time visualization tools and refined bottleneck analysis techniques could further support educational administrators in identifying and mitigating delays or inefficiencies within curricula. Cross-institutional comparisons, facilitated by standardized methodologies and metrics, would enable universities to benchmark and enhance their curricular practices, providing a broader perspective on educational effectiveness and impact. Lastly, detecting concept drift—or significant changes in student progression patterns over time—could help institutions continuously adapt their curricula to evolving educational needs and student demographics.

\subsection*{Funding Information}
No funding has been received for this review.

\subsection*{Conflict of Interest Statement}
The authors have declared no conflicts of interest for this article.

\subsection*{Data Availability Statement}
The list of selected studies is provided as supplementary material at \url{http://dx.doi.org/10.5281/zenodo.14253621}

\subsection*{ORCID}
Daniel Calegari \orcidlinkf{0000-0001-9506-7619}\\
Andrea Delgado \orcidlinkf{0000-0003-4749-9366}


\end{document}